\documentclass[%
reprint,
superscriptaddress,
amsmath,amssymb,
aps,
prl,
]{revtex4-2}

\usepackage{graphicx}% Include figure files
\usepackage{dcolumn}% Align table columns on decimal point
\usepackage{bm}% bold math
\usepackage{gensymb}
\usepackage{subfloat}
\usepackage{xcolor}
\usepackage[normalem]{ulem}
\usepackage{subfigure}
\usepackage{hyperref}% add hypertext capabilities
\usepackage{xcolor}
\definecolor{Blue}{rgb}{0,0,1}
\definecolor{Red}{rgb}{1,0,0}
\definecolor{Green}{rgb}{0,0.52,0.0}
\definecolor{orange}{rgb}{1,0.5,0}
\definecolor{orange2}{rgb}{1,0.5,0.5}

\begin{document}
	
	\preprint{APS/123-QED}
	
	\title{Critical role of electron-phonon interactions in determining the 
	relative stability of Boron Nitride polymorphs}% Force line breaks with \\
	
	\author{Shilpa Paul}%
	%\email{First.Author@institution.edu}
	\affiliation{%
		Department of Metallurgical Engineering and Materials Science, Indian Institute of Technology Bombay, Mumbai 400076, India%\textbackslash\textbackslash
	}%
	
	\author{M. P. Gururajan}%
	\affiliation{%
		Department of Metallurgical Engineering and Materials Science, Indian Institute of Technology Bombay, Mumbai 400076, India%\textbackslash\textbackslash
	}%

	\author{Amrita Bhattacharya}%
	\affiliation{%
		Department of Metallurgical Engineering and Materials Science, Indian Institute of Technology Bombay, Mumbai 400076, India%\textbackslash\textbackslash
	}%
	\author{T. R. S. Prasanna}%
	\email{prasanna@iitb.ac.in}
	\affiliation{%
		Department of Metallurgical Engineering and Materials Science, Indian Institute of Technology Bombay, Mumbai 400076, India%\textbackslash\textbackslash
	}%
	
	\date{\today}% It is always \today, today,
	
\begin{abstract}
Despite several first principles studies, the relative stability of BN 
polymorphs remains controversial. The stable polymorph varies between the cubic (c-BN) and hexagonal (h-BN) depending on the van der Waals 
(vdW) dispersion approximation used. These studies are unable to
explain the main experimental results, c-BN is stable, the relative stability order and the large energy difference between h-BN and c-BN (greater than 150 meV/formula unit). In this study, we introduce 
contributions from electron-phonon interactions (EPI) to the total energy of BN polymorphs. This clearly establishes c-BN is the stable polymorph irrespective of the vdW approximation. Only by including EPI contributions do the \textit{ab initio} results match, for the first time, the main experimental results mentioned above. The EPI contribution to the total energy is strongly sensitive to chemical bonding (approximately twice in $sp^2$-bonded layered over $sp^3$-bonded polymorphs) and to crystal structure. The crucial role of EPI contributions is seen in $sp^2$-bonded layered BN polymorphs where it is greater than the vdW contribution. Given that h-BN is a prototype layered material, in bulk or 2D form, our results have a broader 
relevance, that is, including EPI correction, along with vdW approximation, is vital for the study of energetics in layered materials. 
 
\end{abstract}
	
	\maketitle
	
	%\tableofcontents
	
	Boron Nitride (BN) has four main polymorphs - the $sp^3$-bonded cubic 
	(c-BN) and wurtzite (w-BN) and the $sp^2$-bonded layered hexagonal (h-BN) 
	and rhombohedral (r-BN) with weak van der Waals (vdW) interlayer bonding. 
	Among them, c-BN has been of great interest in structural applications 
	due to its hardness and stabilty \cite{Monteiro2013cubic}.  Recently, there 
	has been an enormous interest in $sp^2$-bonded 2-dimensional (2D) h-BN due 
	to its importance in electronic and photonic applications 
	\cite{Caldwell2019photonics, Roy2021structure}. Because of the importance 
	of BN, there have been several studies to determine the stable structure 
	under ambient conditions. However, controversy exists in both experimental 
	and computational studies between c-BN and h-BN as the stable structure 
	under ambient conditions.	

	Based on high pressure and high temperature (HPHT) studies, Bundy and 
	Wentorf \cite{Bundy1963direct} first suggested that c-BN is the stable 
	phase under ambient conditions. Subsequently, Corrigan and Bundy 
	\cite{Corrigan1975direct} revisited the problem and suggested that 
	h-BN is stable. Later, Solozhenko and co-workers  	
	\cite{Solozhenko1995boron, 	Gavrichev1993low} concluded that c-BN is 
	stable.  Fukunaga \cite{Fukunaga2000equilibrium} concluded that h-BN 
	is stable and c-BN appears stable due to kinetic factors. Soon thereafter, 
	Will \textit{et al.} \cite{Will2000new} concluded that, even after taking kinetic 
	factors into account, c-BN is the stable polymorph under ambient 
	conditions. Clearly, the HPHT experiments are unlikely to lead to 
	consensus on the stable phase due to the slow kinetics of c-BN 
	$\rightarrow$ h-BN transformation.

	However, Differential Thermal Analysis (DTA) studies  
	\cite{Solozhenko1992thermoanalytical, Sachdev1997investigation} 
	show that the transformation from c-BN is endothermic implying that it is 
	the stable polymorph. For comparison, if 
	c-BN was metastable, the above DTA studies would have indicated exothermic 
	transition. The clear contrast in DTA studies 
	makes the results very credible. Thus, the two independent DTA studies 
	\cite{Solozhenko1992thermoanalytical, Sachdev1997investigation}  clearly 
	establish that c-BN is the stable polymorph. 
	
	Solozhenko \cite{Solozhenko1995boron} and Gavrichev \textit{et al.} 
	\cite{Gavrichev1993low} have reported thermodynamic data (enthalpies of 
	formation, enthalpy differences between 298 K and 0 K) for the BN 
	polymorphs. From their data, we obtain the relative stability order (at 0 K)
	and energy differences with respect to c-BN (in units of kJ/mol//meV/f.u.) 
	as c-BN $>$ w-BN (3.5//36.3) $>$ h-BN (15.1//156.5) $>$ r-BN (17.9//185.5). See Supplementary Information for details.

	In \textit{ab initio} density functional theory (DFT) studies on BN polymorphs, the inclusion of van der Waals (vdW) 
	dispersion approximation is essential because of the weak vdW interlayer bonding in the  $sp^2$-bonded layered polymorphs. The stable structure 
	obtained in \textit{ab initio} studies varies depending on the choice of the vdW dispersion 
	approximation used. For example, Cazorla \textit{et al.} \cite{Cazorla2019} have shown 
	that c-BN is stable when the random phase approximation (RPA), many-body 
	dispersion (MBD), fractionally ionic (FI) and the DFT-D3(BJ) dispersion 
	approximations are used. In contrast, studies using coupled cluster theory (CCSD), 
	RPA with exchange (RPAx), fixed-node diffusion Monte Carlo (FNDMC) and 
	DFT-D3 dispersion approximations have shown that layered h-BN (or r-BN) is the 
	stable polymorph \cite{Gruber2018ab, Hellgren2021random, 
	Nikaido2022diffusion}.
	
	However, even the \textit{ab initio} studies that report c-BN as the stable 
	polymorph do not address the serious discrepancies with the experimental 
	results. For example, Cazorla \textit{et al.} \cite{Cazorla2019} have 
	reported the stability order to be c-BN $>$ h-BN $>$ r-BN  $>$ w-BN with 
	h-BN, r-BN $\sim$ 25 meV/f.u. higher in energy than c-BN, whereas the 
	experimental values are $>$ 150 meV/f.u. as discussed above 
	\cite{Solozhenko1995boron, Gavrichev1993low}. Clearly, despite c-BN 
	being the stable polymorph \cite{Cazorla2019}, neither the stability order 
	nor the magnitudes of energy differences match the experimental results 
	\cite{Solozhenko1995boron, Gavrichev1993low}. 	

	In \textit{ab initio} stability studies on BN polymorphs, only the vdW 
	dispersion approximation and the zero-point vibration energy (ZPVE) 
	contributions have been considered. The contribution of electron-phonon 
	interaction (EPI) to the total energy has never been considered. 

	Electron-phonon interactions play an important role in several electronic 
	and optical properties \cite{Giustino2017, Ponce2014a}.  EPI causes the 
	electron eigenenergies to be temperature dependent and given by 
	$E_{n\textbf{k}}(T) = \epsilon_{n\textbf{k}} + \Delta 
	\epsilon_{n\textbf{k}}(T)$, where 
	$\epsilon_{n\textbf{k}}$ is the static lattice eigenenergy for band $n$ and 
	wave vector $\mathbf{k}$ and $\Delta \epsilon_{n\textbf{k}}(T)$ is the EPI 
	correction \cite{Giustino2017, Ponce2014a}. This accounts for the 
	experimentally observed temperature dependence of bandgaps in 
	semiconductors and insulators. Based on the Allen-Heine theory to calculate 
	the EPI contributions \cite{Allen1976, 
	Allen1978, Allen1980, Allen1983, Allen1994}, the temperature dependence 
	of band gaps and band structures for several semiconductors have been 
	calculated \cite{Marini2008, Giustino2010, Antonius2014, Ponce2014b, 
	Ponce2015, Antonius2015, Patrick2014, Zacharias2016, Friedrich2015, 
	Nery2018, Tutchton2018, Querales2019, Miglio2020}.

	The EPI changes the eigenenergies even at 0 K due to  
	zero-point renormalization (ZPR). It follows that the total energy will 
	also be	altered. Allen has developed a general formalism  \cite{Allen2020}, including the recent second Erratum  \cite{Allen2022erratum}, that goes beyond the Quasi-Harmonic Approximation (QHA) and includes contributions to the free energy	from quasiparticle interactions (electron-phonon, 
	phonon-phonon etc.). We have shown that, for semiconductors and insulators, the EPI contributions to the free energies of electrons and phonons are constant (zero-point energy) terms \cite{Varma2022}.

The EPI contribution to the total energy at 0 K, $\Delta E_{EP} (V,0)$, is given by \cite{Allen2022erratum}
\begin{multline}
	\Delta E_{EP} (V,0) = \\
\sum_k \left[\langle k 
|V^{(2)}|k\rangle + \sum_Q \frac{|\langle k 
|V^{(1}|k+Q\rangle|^2}{\epsilon_k - 
\epsilon_{k+Q}}(1-f_{k+Q}) \right]\;f_k
	\label{EPI-zero-K-Allen}
\end{multline}

	 The ZPR of VBM($\Gamma$) in c-BN and h-BN  differ by more than 100 meV \cite{Tutchton2018}. This empirically suggests that the EPI contribution to the total energy differences is also likely to be greater than 100 meV/f.u. \cite{Varma2022}. Hence, it is 
	 essential to study the EPI contribution to the total energy differences in BN polymorphs.

	In this study, we calculate the EPI correction to the total energy for 
	six ($sp^3$-bonded and $sp^2$-bonded layered) BN polymorphs. For the first 
	time, we are able to mostly account for the main experimental results 
	\cite{Solozhenko1995boron, Gavrichev1993low} including the stability 
	order and the large ($>$ 150 meV/f.u.) energy difference between c-BN and 
	h-BN. Our results highlight the critical role of EPI contributions, 
	especially in the study of energetics of $sp^2$-bonded layered BN 
	polymorphs where it is greater than the vdW contribution. Our results imply 
	that studies of energetics in layered materials must include both, vdW 
	dispersion and EPI corrections, and not just the former. 

	The terms in the square bracket in Eq. \ref{EPI-zero-K-Allen} differ from 
	the expression for the EPI correction to the eigenenergy,  $\Delta 
	\epsilon_{n\textbf{k}}$, only by the presence of the $1-f_{k+Q}$ factor 
	in the second term \cite{Allen2022erratum}. At present, the available 
	software codes can calculate $\Delta \epsilon_{n\textbf{k}}$. That is, Eq. 
	\ref{EPI-zero-K-Allen} cannot be calculated at present. Therefore, we 
	approximate Eq. \ref{EPI-zero-K-Allen} as the sum over EPI correction to 
	the eigenenergies given by
	
	\begin{multline}
	\Delta E_{EP} (V,0) \approx \Delta E^{ep}_{av}(V,0) = \\ \sum^{occ}_K 
	\Delta \epsilon_{n\textbf{k}}(V,0)  =  2 
	\sum_{n,\textbf{k}}^{occ,IBZ} w_\textbf{k}\Delta \epsilon_{n\textbf{k}} 
	(V,0) 
	\label{EPI-approx-Abinit} 
	\end{multline}

	The summation is over all the occupied states in the irreducible Brillouin 
	Zone (IBZ) that contribute to the band-structure energy with weight 
	$w_\textbf{k}$. The error due to the neglect of the $1-f_{k+Q}$ factor 
	affects only part of the Fan-Migdal self-energy term and the Debye-Waller 
	term is unaffected \cite{Allen2022erratum}. 

	However, the main interest is in the energy differences between polymorphs. 
	In this case, due to the neglect of the $1-f_{k+Q}$ factor in all 
	polymorphs, the cancellation of errors is likely to reduce the error. 
	Thus, the trends observed using Eq. \ref{EPI-approx-Abinit} are likely 
	to be valid even when the correct expression, Eq. 
	\ref{EPI-zero-K-Allen}, is used. Further justification follows from the good match with the experimental data, for BN polymorphs, for the first time, as seen below.

	For all computations, the ABINIT software package \cite{Gonze2009, 
	Gonze2016} was used. The ONCV \cite{HamannONCV} pseudopotential with PBE \cite{PerdewPBE} exchange-correlation functional was used. To check for consistency, studies using the GGA.fhi pseudopotential were also performed. (See Supplementary Information.)

	The BN polymorphs studied are i) the $sp^3$-bonded c-BN and  w-BN,   
	and ii) the $sp^2$-bonded two-layered hexagonal polymorphs,  h-BN 
	(BN-Aa) and BN-Ab and the three-layered r-BN (BN-ABC) and BN-ABb. The 
	notations are taken from Luo \textit{et al.} \cite{Luo2017new}. The 
	$sp^2$-bonded polymorphs were selected because they are elastically and 
	dynamically stable \cite{Luo2017new}.

	The DFT-D3 \cite{Grimme2010consistent} and DFT-D3(BJ) \cite{Grimme2011} 
	dispersion approximations were used to calculate 
	the total energies and lattice parameters of all BN 
	polymorphs. The EPI contributions for each eigenstate was calculated using 
	the ABINIT module, temperature dependence of the electronic structure 
	(TDepES) \cite{Abinit}. The computational details are given in the 
	Supplementary Information.

		\begin{table}[h]
			\caption{\label{table1}%
				Lattice parameters, ZPR of VBM($\Gamma$), CBM($\Gamma$) and 
				energy stability of BN polymorphs relative to c-BN for DFT-D3 
				and DFT-D3(BJ) vdW approximations.}
			%\begin{ruledtabular}
			%\begin{tabular}{p{2.7cm}p{2.9cm}p{2.9cm}p{2.9cm}}
			\begin{tabular}{p{1.8cm} p{1.7cm}  p{1cm}  p{1cm} c c}
			\hline\hline
				Polytype & a, c&\multicolumn{2}{c}{ZPR} & 
				\multicolumn{2}{c}{$\Delta$E}\\
				& & VB($\Gamma$)& CB($\Gamma$)& D3& D3(BJ) \\
				&(Bohr) &(meV) &(meV)
				& \multicolumn{2}{c}{(meV/f.u.)} \\
				\hline
				c-BN-D3 & 6.82 & 174.8 &-256.9 & 0&{}\\
				c-BN-BJ & 6.81 & 174.6 &-255.9 & {} & 0 \\
				w-BN-D3 &4.8, 7.95 & 151.9 &-123.6 & 36.5&{} \\
				w-BN-BJ & 4.8, 7.94 & 151.7  &-124.2 & {} & 35.4 \\
				\hline
				h-BN-D3 & 4.73, 12.64 & 262.3 &-150.2 & -29&{} \\
				h-BN-BJ &4.73, 12.48 & 262.8  &-144.7 & {} & 5.5 \\
				BN-Ab-D3 & 4.73, 12.75 & 256.3 &-161.3 & -24.3&{}\\
				BN-Ab-BJ & 4.73, 12.58 & 255.2  &-159.6 & {} & 8.1 \\
				\hline
				r-BN-D3 & 4.73, 18.9 & 253.7 &-154.4 & -29.3&{} \\
				r-BN-BJ & 4.73, 18.69 & 252.8  &-147.3 & {} & 5.2 \\
				BN-ABb-D3 & 4.74, 19.11 & 257.7 &-199.2 & -27.2& {} \\
				BN-ABb-BJ & 4.73, 18.78 & 256.4  &-200.7 & {} & 6.7 \\
				\hline\hline
				
			\end{tabular}
			%\end{ruledtabular}
		\end{table}

	Table \ref{table1} gives the lattice parameters, ZPR for VBM and CBM at 
	$\Gamma$-point and the relative stability of BN polymorphs, with 
	respect to c-BN, for the DFT-D3 and DFT-D3(BJ) dispersion approximations. 
	Our results are similar to literature values \cite{Cazorla2019, 
	Nikaido2022diffusion}. 

	The layered polymorphs, r-BN and h-BN, are almost degenerate for both 
	DFT-D3 and DFT-D3(BJ) 
	approximations. They are also the stable structures for DFT-D3 
	approximation while c-BN is stable for the DFT-D3(BJ) approximation, 
	similar to literature reports \cite{Cazorla2019}. 	

	The ZPR for c-BN at VB($\Gamma$) is similar to that of Ponce \textit{et 
	al.} \cite{Ponce2015} though the value of Tutchton \textit{et al.} 
	\cite{Tutchton2018} is different.  For h-BN, our 
	value is similar to that of Tutchton \textit{et al.} \cite{Tutchton2018}. 
	(Additional ZPR values at VBM and CBM are reported in Supplementary 
	Information.)
	
	An advantage of the DFT-D\textit{n} (D2, D3, D3(BJ) etc.) methods is that they are compatible with the EPI calculation method used in the present study \cite{Troeye2016}. Thus, by calculating the EPI corrections at the 
	DFT-D\textit{n} lattice parameters both, vdW and EPI contributions, are 
	incorporated \cite{Troeye2016,Tutchton2018}.

\begin{figure}[h]
		\includegraphics[scale=0.28]{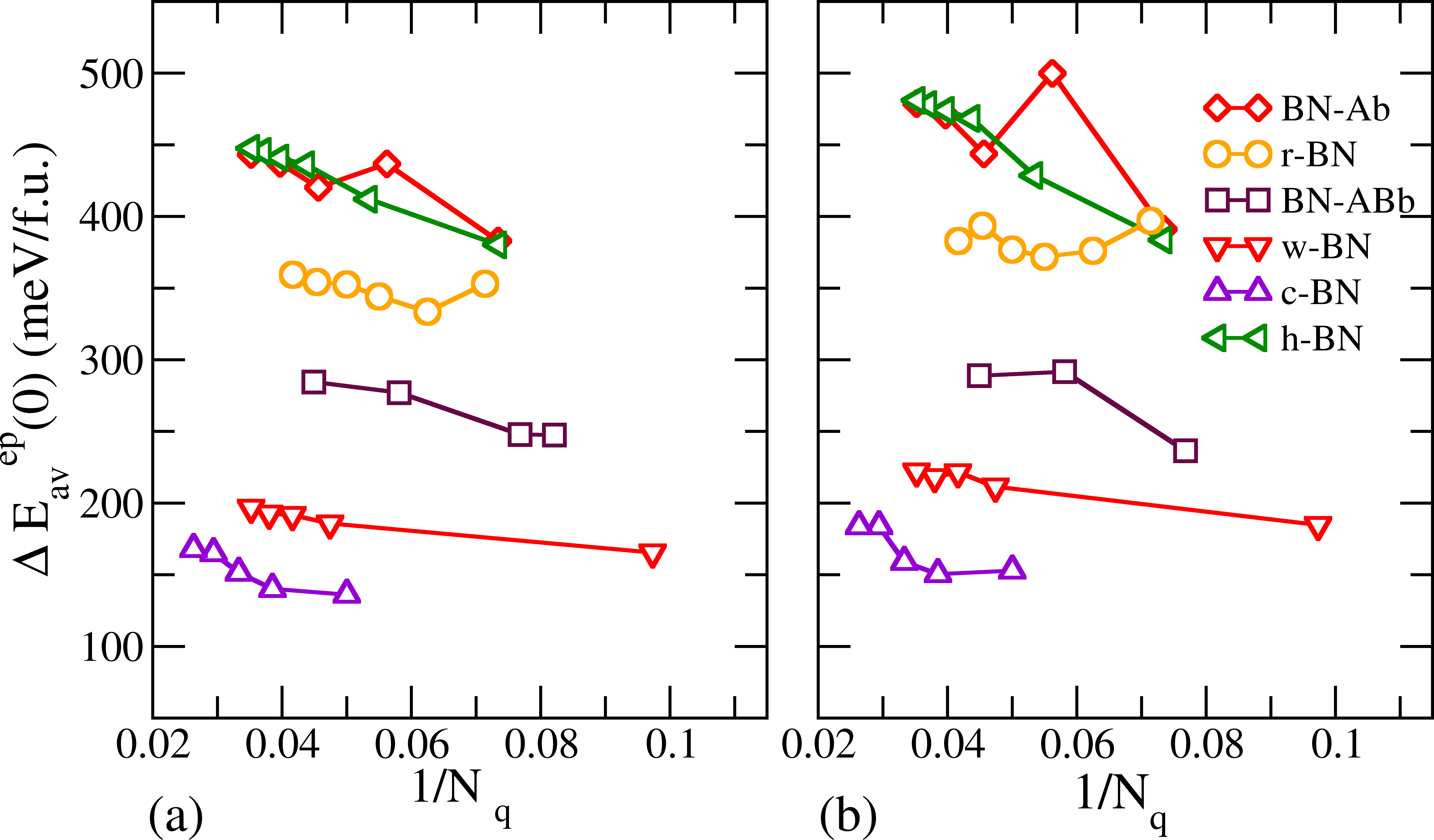}
		\caption{\label{fig1} Convergence of EPI correction to total energy 
		with $\mathbf{q}$-point grid density (non-adiabatic approximation) for 
		smearing parameters: (a) 100 meV and (b) 50 meV, for BN polymorphs for 
		DFT-D3(BJ) lattice parameters. }
	\end{figure}

	Fig. \ref{fig1} shows the convergence of the EPI correction to the total 
	energy due to zero-point renormalization (0 K), $\Delta E^{ep}_{av}(0)$, 
	with $1/N_q$ where $N_q^3$ is the total number of $\mathbf{q}$-points in 
	the Brillouin Zone (BZ). 

	A clear trend is observed with $\Delta E^{ep}_{av}(0)$ smallest for the 
	$sp^3$-bonded polymorphs (c-BN, w-BN), followed by $sp^2$-bonded 
	three-layered polymorphs (BN-ABb, r-BN) and the highest for the 
	$sp^2$-bonded two-layered BN polymorphs (h-BN, BN-Ab). Clearly, 
	$sp^2$-bonding leads to much higher $\Delta E^{ep}_{av}(0)$ than 
	$sp^3$-bonding in BN polymorphs. Further, the magnitude of the differences 
	in $\Delta E^{ep}_{av}(0)$ between the $sp^3$-bonded c-BN and the 
	$sp^2$-bonded layered BN polymorphs are in the range of 110-300 
	meV/f.u.  Within $sp^2$-bonded BN polymorphs, three-layered polymorphs have 
	lower $\Delta E^{ep}_{av}(0)$ than two-layered polymorphs. Among them, 
	$\Delta E^{ep}_{av}(0)$ is 	significantly different between r-BN (1 
	f.u./u.c.) and ABb (3 f.u./u.c.), 	which suggests significant crystal 
	structure influence. However,  $\Delta E^{ep}_{av}(0)$ for the two-layered 
	BN polymorphs (2 f.u./u.c) are 	almost degenerate indicating 
	little crystal structure influence among them. In summary, $\Delta 
	E^{ep}_{av}(0)$ is primarily influenced by bonding ($sp^2$ over $sp^3$) 
	and then by crystal structure.
	
  	For DFT-D3 lattice parameters, the EPI calculations were performed only 
	for the highest $\textbf{q}$-point grid (the smallest $1/N_q$ in 
	Fig.\ref{fig1}). The trends are similar to those for DFT-D3(BJ) lattice parameters (Table \ref{table2}). 
	
	Table \ref{table2} lists the vdW and EPI contributions to the total energy 
	for BN polymorphs. Both depend primarily on the bonding type. The vdW 
	contribution is higher for $sp^3$-bonded polymorphs over $sp^2$-bonded 
	polymorphs. However, within each bonding type it is very weakly dependent on crystal 
	structure. In contrast, the EPI contribution is higher ($\sim$ 2 times) for $sp^2$-bonded compared to 
	$sp^3$-bonded polymorphs. The EPI contribution is also dependent on the 
	crystal structure. Of particular interest are the contributions for the  $sp^2$-bonded layered BN polymorphs where the EPI contribution is 
	$\sim$ 1.5 - 2 times greater than the vdW contribution. Eventhough, currently only the vdW contribution is  included, our results clearly imply that EPI contribution must also be included in the studies on the energetics of layered BN polymorphs.

	\begin{table}[h]
		\caption{\label{table2}%
			Comparison of the vdW and EPI contributions (meV/f.u.) to total 
			energies of BN polymorphs for D3 and D3(BJ) approximations. }
		\begin{tabular}{p{1.7cm} c c c c}
			\hline\hline
			Polytype & \multicolumn{2}{c}{vdW} &  \multicolumn{2}{c}{EPI} \\
			& D3(BJ) & D3 & D3(BJ) & D3 \\
			
			\hline
			c-BN & -416.8 & -316.1 & 168.1 &  168.4\\
			w-BN &  -420.8 & -316.5 & 196.8 & 196.3\\
			\hline
			h-BN & -284.1 & -212.4 & 447.6 & 448.5 \\
			BN-Ab &  -281.4 & -208.4 & 443.3 & 445.2 \\
			\hline
			r-BN & -284.7 & -213.6 & 359.4 & 351.6 \\
			BN-ABb & -283.2 & -209.2 & 284.5 & 284.4 \\
			\hline\hline
			
		\end{tabular}
		
	\end{table}

	The EPI contribution to the phonon free energy consists of two components - 
	adiabatic and non-adiabatic \cite{Giustino2017, Heid201312}. The 
	non-adiabatic component can be neglected for large band-gap semiconductors 
	and insulators because the band-gap is much greater than the phonon energy 
	\cite{Giustino2017, Calandra2010adiabatic}. The adiabatic component is 
	already included in a density functional perturbation theory (DFPT) 
	calculation \cite{Allen2020, Giustino2017, Heid201312}. The phonon spectra	
	for BN polymorphs using DFPT and finite difference method (FDM) are 
	 similar \cite{Li2019continuity, Karch1996lattice, Karch1997ab, 
	 Ohba2001first, Kern1999ab, Yu2003ab} implying that both methods, 
	 explicitly and implicitly, include the adiabatic EPI contribution 
	 \cite{Allen2022review}. 

\begin{figure}[h]
		\includegraphics[scale=0.28]{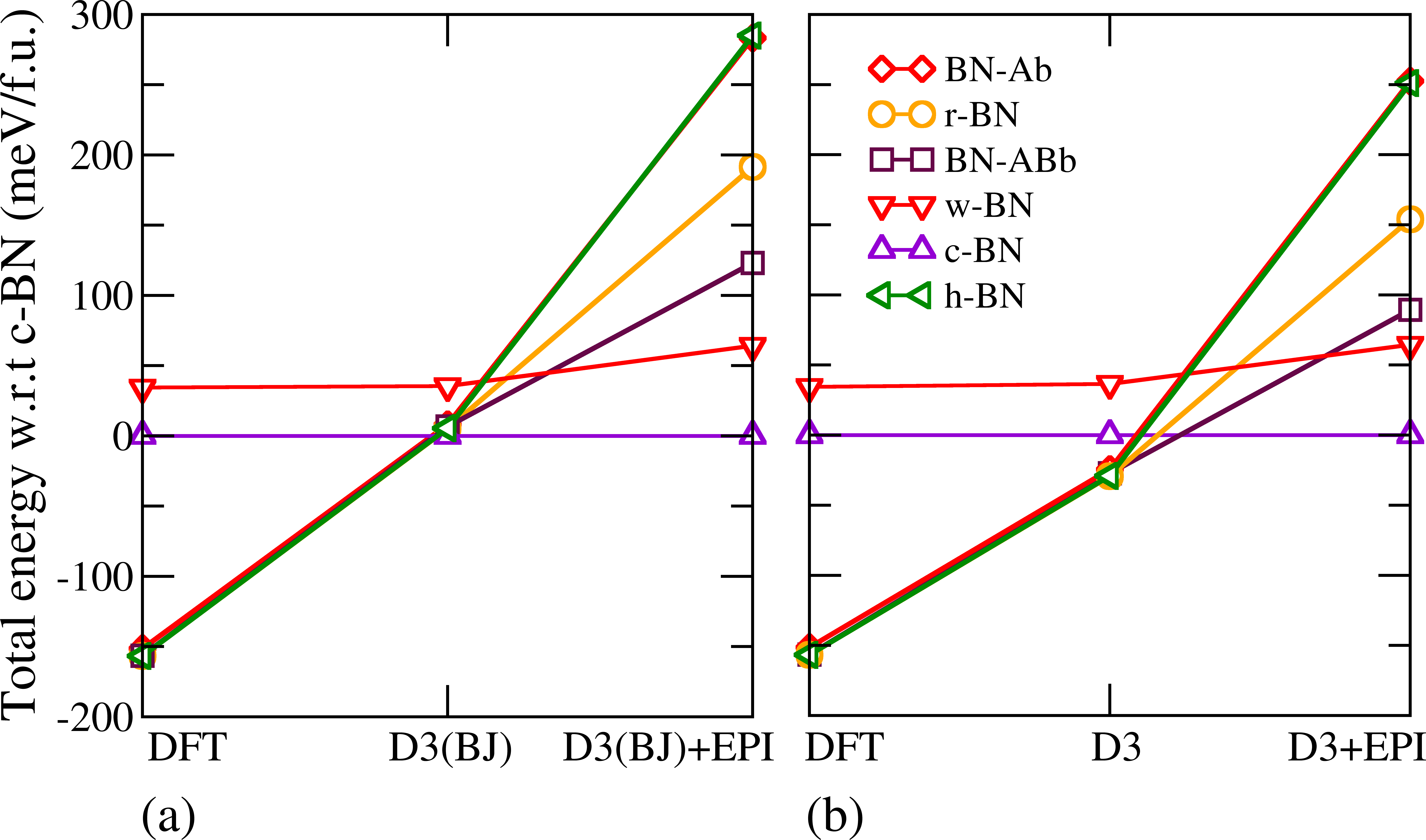}
		\caption{\label{fig2} Relative stability of BN polymorphs with respect 
		to c-BN for (a) DFT, DFT-D3(BJ) and DFT-D3(BJ) + EPI and (b) DFT, 
		DFT-D3 and DFT-D3 + EPI conditions. }
	\end{figure}

	Fig. \ref{fig2} shows the final stablity order of the BN polymorphs after 
	including the vdW and EPI corrections. The ZPVE contributions 
	(Supplementary Information) are similar to literature values 
	\cite{Cazorla2019}. They vary among BN polymorphs by $<$ 10 mev/f.u. and 
	are unlikely to alter the stability order. Under DFT conditions (without 
	vdW corrections), c-BN is metastable and 	h-BN is stable consistent with 
	literature \cite{Cazorla2019, Nikaido2022diffusion}. The DFT-D3 
	approximation stabilizes h-BN while DFT-D3(BJ) stabilizes c-BN, consistent 
	with literature \cite{Cazorla2019}. 

	However, when the EPI corrections are included, c-BN is the stable 
	polymorph irrespective of whether DFT-D3 or DFT-D3(BJ) approximation is used. Further, the stability order of BN polymorphs is 
	similar in both the vdW approximations  - c-BN $>$ w-BN $>$ h-BN(ABb) $>$ 
	r-BN $>$ h-BN, 	h-BN(Ab) and h-BN(AC), where all the two-layered polymorphs are almost degenerate.
	
	In Table \ref{table3} we compare our results with computational 
	\cite{Cazorla2019,Nikaido2022diffusion} and  
	experimental results \cite{Solozhenko1995boron, Gavrichev1993low} for the four major BN polymorphs. Because the DFT-D3 and DFT-D3(BJ) results are similar to FNDMC, RPAx, CCSD  and RPA, FI, MBD results respectively 
	\cite{Nikaido2022diffusion, Cazorla2019}, we can consider them to be 
	representative of all dispersion approximations for BN total energy studies. Table \ref{table3} shows that neither RPA and DFT-D3(BJ) (c-BN stable) results nor FNDMC and DFT-D3 results (h-BN stable) are consistent with the experimental energy differences.

		\begin{table}[h]
		\caption{\label{table3}%
			Total energy differences  with respect to c-BN of four BN polymorphs obtained in the present study compared with reported
			experimental and computational results.}
		\begin{tabular}{p{2.7cm} c c c c}
		\hline\hline
			$\Delta E$ (meV/f.u.) & c-BN &w-BN &r-BN&h-BN\\
			\hline
			Experimental \cite{Solozhenko1995boron, Gavrichev1993low} & 0 & 36 
			$\pm$ 32 & 185  $\pm$ 42&155 $\pm$ 31\\
			\hline
			RPA \cite{Cazorla2019}&0 &38.8 & 29.5 & 25.6\\
			D3(BJ)*&0 &35.4 & 5.2 & 5.5\\
			D3(BJ)+EPI*&0 &64.2 & 191.5 & 285.1\\
			\hline
			FNDMC \cite{Nikaido2022diffusion}&0 &44.6 & -21 & -36.3\\
			D3*&0 &36.6 & -29.3 & -29\\
			D3+EPI*&0 &65.3 & 157.1 & 250.6\\
			\hline\hline
			*this study
		\end{tabular}
	\end{table}

	It is only when EPI contributions are 	included that we obtain results 
	that are in reasonable agreement with the experimental results. Firstly, 
	including EPI corrections leads to c-BN as the stable polymorph 
	irrespective of the vdW dispersion approximation used. Secondly, w-BN is 
	the lowest energy metastable polymorph similar to the experimental results. For the $sp^2$-bonded 
	layered polymorphs, the energy difference for r-BN matches well with the experimental value. However, for h-BN, our results are an overestimation compared to the experimental value. Overall, the $sp^2$-bonded h-BN and r-BN are higher in energy by $>$ 150 meV/f.u. and such large energy differences have never been obtained in any previous \textit{ab initio} study. Only the inclusion of EPI contributions enables, for the first time, the \textit{ab initio} 
	results to mostly match the experimental results. Hence, we can attribute
	the experimentally observed $>$ 150 meV/f.u. total energy differences  
	between c-BN and h-BN, r-BN to be primarily due to differences in 
	EPI contributions. The critical role of EPI contributions in matching 
	\textit{ab initio} results with experimental results is clear.
	
	The agreement with experimental energy differences indicates that the 
	$1-f_{k+Q}$ factor, ignored in Eq.\ref{EPI-approx-Abinit}, is clearly not 
	the dominant term and does not significantly affect the magnitude of 
	the energy differences.

	Our results have important implications. Hexagonal h-BN and graphene are 
	among the most investigated layered materials, in bulk and 2D forms 
	\cite{Bjorkman2012van, Bjorkman2014testing, Lebedeva2017comparison, 
	Tawfik2018evaluation}. In	general, the reduced dimensionality leads to 
	increased EPI strength in 2D materials \cite{Tutchton2018, 
	Giustino2021exciton}. Thus, Table \ref{table2}) implies a 
	distinct possibility that the EPI contributions for 2D h-BN and graphene 
	are greater than the vdW contributions and hence, must be included in the 
	study of energetics of the same.

	Furthermore, h-BN and graphene, in bulk and 2D forms, are the prototype 
	materials for the study of the accuracy of various vdW dispersion 
	approximations in layered materials \cite{Bjorkman2012van, 
	Bjorkman2014testing, Lebedeva2017comparison, 
	Tawfik2018evaluation}. For such studies, the benchmark reference total 
	energy is obtained from either RPA or diffusion monte carlo (DMC) 
	studies.  However, as seen from Table \ref{table3}, the RPA and 
	FNDMC results are not consistent with experimental values. It is critical 
	to include EPI contributions to obtain \textit{ab initio} results 
	consistent with experimental results. Thus, our results imply that for 
	layered materials, in bulk and 2D forms, the accuracy of vdW dispersion 
	approximations cannot be studied in isolation and the benchmark reference 
	total energy must also include EPI contributions. 

	As an initial method to study the energetics of $sp^2$-bonded layered 
	materials, in bulk or 2D form, we propose the DFT-D\textit{n} plus EPI 
	combination because they are compatible and the EPI calculations must be 
	performed at the DFT-D\textit{n} lattice parameters \cite{Troeye2016}. 
	However, if more sophisticated dispersion approximations are to be used 
	e.g. MBD, RPA, DMC etc., then compatible EPI calculation methods need to be 
	developed.

	In conclusion, we resolve the long-standing controversy about the 
	\textit{ab initio} relative	stablility order of BN polymorphs and its 
	disagreement with experimental 
	results. To do so, we have introduced a new term in the total energy, from 
	EPI contribution. This clearly establishes c-BN as the stable structure 
	irrespective of the vdW dispersion approximation used. By 
	including EPI correction to the total energy, for the first	time, we are 
	able to mostly match the experimental results, including the relative 
	stability order and also the magnitudes of energy differences. The EPI 
	contribution to the total energy is most sensitive to chemical bonding 
	($sp^2$ over $sp^3$) followed by crystal structure. The experimentally 
	observed large ($>$ 150	meV/formula unit) difference in total energy 
	between h-BN (and r-BN) and c-BN is primarily due to the differences in EPI 
	contributions for the $sp^2$ and $sp^3$-bonded BN polymorphs. Since h-BN is 
	a prototype layered material, our results suggest that it is critical to 
	include EPI corrections, along with vdW dispersion corrections, in studies 
	on the energetics of all layered materials, in bulk or 2D form. 

\begin{acknowledgments}
The authors are very thankful to Prof. P. B. Allen for sharing a draft of second Erratum \cite{Allen2022erratum} for Ref. \cite{Allen2020}. We also thank the ``Spacetime'' HPC facilities at IIT Bombay for computational support.
\end{acknowledgments}

	\bibliography{references.bib}% Produces the bibliography via BibTeX.

\end{document}